\documentclass[10pt, conference, letterpaper]{IEEEtran}

\ifCLASSOPTIONcompsoc
  \usepackage[nocompress]{cite}
\else
  \usepackage{cite}
\fi
\usepackage{cite,url}
\usepackage{booktabs}
\usepackage{amsmath,amssymb,bm,dsfont,bbm,mathtools}
\usepackage{algorithmic}
\usepackage{amsthm,nccmath,lipsum,cuted}
\usepackage[dvips]{color}
\usepackage{graphicx}
\usepackage[lined,boxed,commentsnumbered, ruled]{algorithm2e}
\usepackage{subcaption}
\usepackage{cleveref}
\usepackage{amsfonts}
\usepackage{ragged2e}

\newtheorem{proposition}{Proposition}
\usepackage{graphicx}
\usepackage{verbatim}
\usepackage{comment}

\usepackage[footnote,draft,danish,silent,nomargin]{fixme}

\def\P{{\mathbb P}}  

\graphicspath{{./figs/}}

\makeatletter
\newcommand{\pushright}[1]{\ifmeasuring@#1\else\omit\hfill$\displaystyle#1$\fi\ignorespaces}
\newcommand{\mathleft}{\@fleqntrue\@mathmargin0pt}
\newcommand{\mathcenter}{\@fleqnfalse}
\makeatother
\begin{document}

\title{Detecting State Transitions of a Markov Source: Sampling Frequency and Age Trade-off}
\author{Jaya Prakash Champati, Mikael Skoglund, and James Gross\\
Information Science and Engineering, EECS, KTH Royal Institute of Technology, Stockholm, Sweden\\
E-mail: $\{$jpra, skoglund, jamesgr$\}$@kth.se    
}
\maketitle

\begin{abstract}
We consider a finite-state Discrete-Time Markov Chain (DTMC) source that can be sampled for detecting the events when the DTMC transits to a new state. Our goal is to study the trade-off between sampling frequency and staleness in detecting the events. We argue that, for the problem at hand, using Age of Information (AoI) for quantifying the staleness of a sample is conservative and therefore, introduce  \textit{age penalty} for this purpose. We study two optimization problems: minimize average age penalty subject to an average sampling frequency constraint, and minimize average sampling frequency subject to an average age penalty constraint; both are Constrained Markov Decision Problems. We solve them using linear programming approach and compute Markov policies that are optimal among all causal policies. Our numerical results demonstrate that the computed Markov policies not only outperform optimal periodic sampling policies, but also achieve sampling frequencies close to or lower than that of an optimal clairvoyant (non-causal) sampling policy, if a small age penalty is allowed.
\end{abstract}

\section{Introduction}
Detecting the occurrence of an event when monitoring an information source or a process of interest is essential to applications from varied domains that include control and information systems. In a control system, for instance, a sensor samples a process for detecting an event where the state of the process exceeds a certain threshold value. In World Wide Web, a web crawling application is equipped with the task of downloading remote web pages to a local database (for page ranking/indexing etc.), and is required to detect the events when the remote web page gets updated. 

In practice, it is impossible to know the exact time instant of occurrence of an event unless the source is sampled infinitely often (or in every time slot for discrete-time systems). However, sampling at a higher frequency incurs costs to a system in terms of the energy consumption of a sensor, or the bandwidth usage of the network for transmitting the samples. On the other hand, sampling at a lower frequency results in staleness in detecting an event. Therefore, we are interested in the question: \textit{given the source is sampled in time slot $n$, how to choose the next sampling instant $n + \tau$ such that the conflicting objectives average sampling frequency and average staleness in the event detection are optimized?} In this work, we address this question for an information source modelled using a finite-state DTMC and the events we want to detect are transitions of the DTMC to new states. Even though this setting seems fundamental and is useful in modelling different applications, to the best of our knowledge, the trade-off problems we study have not been tackled in the literature -- see Section~\ref{sec:related} for related works. 


The first step in studying the trade-off between sampling frequency and staleness is to choose an appropriate metric for quantifying the staleness of a sample. For this purpose, one may choose Age of Information (AoI), which has emerged as a relevant performance metric for quantifying staleness of updates at a destination in a communication system. It is defined as the time elapsed since the generation of freshest update available at the destination~\cite{kaul_2011a}. However, we argue that using AoI is conservative for the problem at hand and introduce a staleness metric \textit{age penalty}, which is defined as the time elapsed since the first transition out of the most recently observed state. We then formulate two problems: minimize average age penalty subject to an average sampling frequency constraint, and minimize average sampling frequency subject to an average age penalty constraint. Both the problems are Constrained Markov Decision Problems (CMDPs). We use Linear Programming (LP) approach to solve for optimal Markov policies that are known to be optimal among all causal polices for the problems at hand. In our numerical analysis using a two-state Markov chain  we find that, the optimal policy always provides lower sampling frequency than optimal periodic sampling policy and the gap increases with lower probability of transitions. We also present a comparison of the sampling frequency achieved by the optimal policy with that of the sampling frequency of an optimal clairvoyant (non-causal) sampling policy. 

The rest of the paper is organized as follows. In Section~\ref{sec:model}, we present the system model and formulate the CMDPs. The LP solution approach for both the problems is described in Section~\ref{sec:CMDP}. Numerical analysis using a two-state Markov chain is presented in Section~\ref{sec:numerical}. Related work in presented in Section~\ref{sec:related} and we conclude in Section~\ref{sec:conclusion}.

\section{System Model and Problem Statement} \label{sec:model}
\subsection{Markov Source}
We consider an information source/process that is modelled by an $N$-state DTMC $\{X_n,n \geq 0\}$ where $N < \infty$. We assume that the DTMC is ergodic, i.e., irreducible and aperiodic. Let $S = \{1,2,\ldots,N\}$ denote the set of states. We use $p_{ij}$, for all $i,j \in S$, to denote the one-step probabilities, and the $n$-step transition probabilities are denoted by
\begin{align*}
p^{(n)}_{ij} = \P(X_n = i|X_0 = j), \; \forall i,j \in S.
\end{align*}
Given the one-step probabilities, the $n$-step transition probabilities can be computed using matrix multiplication on the one-step transition probability matrix~\cite{norris_1997}.  Let $\xi_j$ denote the stationary probability of finding the DTMC in state $j$.

\begin{figure}[t]
	\centering
	\includegraphics[scale=.35]{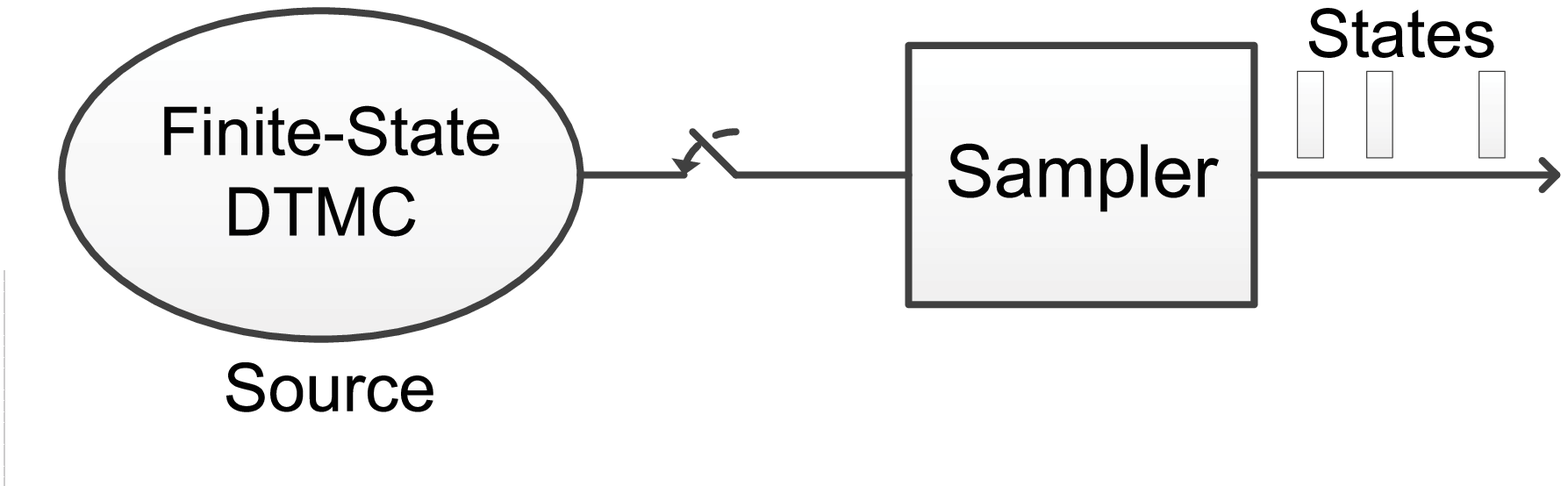}
	\caption{Sampling an information source/process modelled using a DTMC. Each sample reveals the state of the DTMC.}
	\label{fig:system}
\end{figure}

A time slot in the system represents one unit of time of the DTMC and the state transitions occur at the start of a time slot. The state of the DTMC can only be observed by sampling the source; see Figure~\ref{fig:system}. 
Let $T_0 = 0$, $T_1$, $T_2$, $\ldots$ denote the time instants of transitions of the DTMC to new states. We are interested in detecting these transitions at the earliest time possible. Our motivation for studying this problem arises due to its relevance to applications from different domains.
\begin{itemize}
	\item In a control system, the source is a process of interest, and a state transition represents an event where the process exceeds a certain threshold. 
	\item In a web crawling application~\cite{Olsten2010}, the source is a remote web page and the state transitions models the updating events of the website.
\end{itemize}

Clearly, sampling the source at the start of every slot allows us to detect each and every transition of the DTMC. Instead, our aim here is to use lower sampling frequency. This translates to energy savings for a sensor and/or bandwidth savings for transmitting lower number samples to a controller/monitor. In the case of web crawling application, this translates to lower frequency of downloads of the remote web page. However, using lower sampling frequency will result in staleness in detecting a transition and may also miss several transitions. We are thus interested in studying the trade-off between sampling frequency and staleness. Next, we define sampling policies and the age penalty for quantifying staleness.


\subsection{Sampling Policy and Age Penalty}
Assume that $X_0$ is given. A sampling policy $\pi$ specifies the set of sampling instants $\{G_k,k\geq 1\}$, where $G_k$ is the sampling instant of the $k$th sample.  Define $\tau_k = G_k - G_{k-1}$ for all $k \geq 1$, and $G_0 = 0$, then the policy $\pi$ can be equivalently specified by $\{\tau_k,k\geq 1\}$. We assume that $\tau \in Q =  \{1,2,\ldots,M\}$, where $M < \infty$ is the maximum inter-sampling time allowed in the system. Let $\Pi$ denote the set of all causal policies, where a causal policy considers the current and all past observed states and past actions for choosing the current action. In the sequel, we study the following policies.
\begin{enumerate}
	\item \textit{Markov policies:} A Markov policy maps each state to an action with a fixed probability. To be precise, let $j$ be the observed state in the $k$th decision epoch, then under a Markov policy $\tau_k$ is assigned a value $\tau \in Q$ according to a fixed probability distribution $\P^\pi(\tau_k = \tau|j)$. Let $\Pi^\text{MR}$ denote the set of Markov policies.
	\item \textit{Periodic sampling policies:} Under these policies, samples are taken at fixed time intervals $\tau$. With a slight abuse in notation we use $\pi(\tau)$ to denote such a policy. Note that periodic sampling policies are a subclass of Markov policies.
	\item \textit{Optimal clairvoyant sampling policy:} Under this policy, the next transition to a new state is assumed to be known a priori, and thus the source is sampled exactly at the instants when transitions between states occur. Let $\pi^\dagger = \{G^\dagger_k,k \geq 1\}$ denote this policy and $\nu^\dagger$ denote its average sampling frequency. Note that $\pi^\dagger$ is a non-causal policy and we study it for theoretical benchmarking. 
\end{enumerate}

\begin{figure}[t]
	\centering
	\includegraphics[scale=.35]{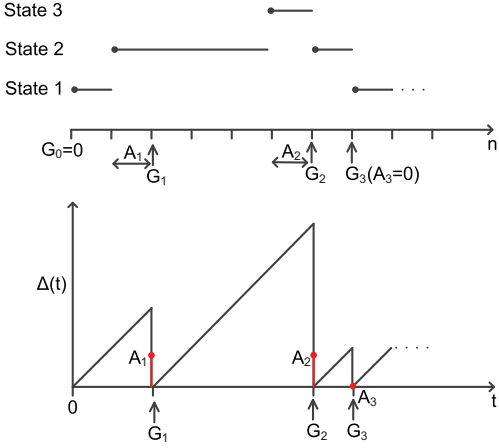}
	\vspace{-.2cm}
	\caption{A sample path of a 3-state Markov chain. AoI and age penalties are depicted for first three sampling instants of a policy with $G_1 = 2, G_2 = 6,$ and $G_3 = 7$.}
	\label{fig:sample_Age}
	\vspace{-.4cm}
\end{figure}

As stated before, sampling the source at the start of every slot allows us to identify each and every transition of the DTMC to a new state and thus staleness of each sample is zero. However, quantifying the staleness of a sample in general is not entirely obvious. This is because, when the sampler samples the source it may find that the DTMC is in the same state or a different state from the previous sample, and even in the former case multiple transitions might have occurred. 
One may consider AoI, denoted by $\Delta(t)$, at the sampler as the staleness metric. It increases linearly between two sampling instants and resets to zero at the sampling instants. However, using this statelessness metric is conservative in this context. To illustrate this, in Figure~\ref{fig:sample_Age} we plot the sample-path of a 3-state DTMC and the resulting AoI. Note that in the duration between the instants $G_1$ and $G_2$, DTMC stays in state $2$ for $3$ time-slots after it was observed by the sampler at $G_1$. Ideally, this should not be accounted for the staleness of the sample at $G_2$, but AoI adds a linear penalty for this duration. 

Using the above insight, we quantify the stateless of a sample $k$ by introducing \textit{age penalty} $A_k$, which is defined as the time elapsed since the first transition out of the state in the $k-1$ sample\footnote{One may also consider including the number of missed transitions in the age penalty and with some effort solve the problem using the same approach in this paper.}.
Under policy $\pi$ the age penalty for the $k$th sample is given by
\begin{align*}
A_k(\pi) = \max\{0,G_k - \min_{n}\{T_n:T_n \geq G_{k-1}\}\}.
\end{align*}  
This entity is illustrated and contrasted with AoI in Figure~\ref{fig:sample_Age}. Under a policy $\pi$, the average age penalty $\mathbb{E}[A(\pi)]$ is given by,
\begin{align*}
\mathbb{E}[A(\pi)] = \limsup_{K \rightarrow \infty} \frac{\mathbb{E}[\sum_{k = 1}^K A_k(\pi)]}{K},
\end{align*}
and the average sampling-interval is given by $\limsup_{K \rightarrow \infty} \frac{\mathbb{E}[\sum_{k=1}^{K}\tau_k]}{K}$, where the expectation is taken with respect to the probability distribution induced by $\pi$ on the sequence of observed states and actions.

\subsection{Optimization problems $\mathcal{P}_1$ and $\mathcal{P}_2$}
We are interested in the following problems. For a given upper bound $\nu \in (0,1]$ on the average sampling frequency, in problem $\mathcal{P}_1$ we aim to minimize the average age penalty which is stated below.
\begin{equation}\label{prob1}
\begin{aligned}
& \underset{ \pi \in \Pi }{\text{minimize}} && \mathbb{E}[A(\pi)]   \\
& \text{s.t.}  &&\limsup_{K \rightarrow \infty} \frac{\mathbb{E}[\sum_{k=1}^{K}\tau_k]}{K} \geq \frac{1}{\nu} \,.
\end{aligned}
\end{equation} 
For a given upper bound $d \geq 0$ on the average age penalty, in problem $\mathcal{P}_2$ we aim to maximize the average sampling-interval which is state below.
\begin{equation}\label{prob2}
\begin{aligned}
& \underset{ \pi \in \Pi }{\text{maximize}} && \limsup_{K \rightarrow \infty} \frac{\mathbb{E}[\sum_{k=1}^{K}\tau_k]}{K}  \\
& \text{s.t.} && \mathbb{E}[A(\pi)] \leq d \,.
\end{aligned}
\end{equation}
Let $\pi^*_1$ and $\pi^*_2$ denote optimal policies for $\mathcal{P}_1$ and $\mathcal{P}_2$, respectively.

\textbf{\textit{Remark:}} For $\mathcal{P}_1$, an optimal periodic sampling policy chooses $\tau = \lceil 1/\nu \rceil$. For $\mathcal{P}_2$ an optimal periodic sampling policy chooses $\tau = d + 1$.

Finally, we define $\tau^{\dagger} = \lceil 1/\nu^{\dagger} \rceil$.

\section{Linear Programming Solution Approach}\label{sec:CMDP}
Both $\mathcal{P}_1$ and $\mathcal{P}_2$ are Constrained Markov Decision Problems (CMDP). A CMDP with finite state and action sets has an optimal policy in the set of Markov policies~\cite{Altman1999}, and it can be efficiently solved using the Linear Programming (LP) approach presented in~\cite{Manne1960}. Therefore, in the following we only need to consider the set of Markov policies. Under Markov policies the  induced stochastic process $\{X_{G_k}, k\geq 1\}$, i.e., the sequence of observed states, is also a DTMC; in the sequel we refer to it as \textit{induced DTMC}.
\subsection{Elements of the CMDP}
The decision epochs in $\mathcal{P}$ are indexed by $k$.
\begin{itemize}
	\item \textbf{State space:} $S = \{1,2,\ldots,N\}$.
	\item \textbf{Action space:} At decision epoch $k$, the next inter-sampling time $\tau_{k+1}$ is chosen from the set $Q =  \{1,2,\ldots,M\}$.
	\item \textbf{Transition probabilities:} The next state $i \in S$ of the induced DTMC only depends on the current observed/sampled state $j$ and the sampling interval $\tau$. To be precise, let $j$ be the state observed in decision epoch $k$, i.e., in time slot $G_k$, then the transition probability of the induced DTMC to state $i$ for any sampling interval $\tau$ is given by
	\begin{align*}
	q_{j\tau i} &= \P(X_{G_{k}+\tau} = i|X_{G_k}=j)\\
	&= \P(X_{\tau} = i|X_0 = j) \\
	&= p^{(\tau)}_{ji}, \; \forall i,j \in S \text{ and }\tau \in Q. 
	\end{align*}
	Further, given $\pi \in \Pi^\text{MR}$, the stead-state probabilities $\lim_{k \rightarrow \infty} \P^\pi(X_{G_k} = j)$ for the induced DTMC can be computed from the following transition probabilities.
	\begin{align}\label{eq:transInducedDTMC}
	&\P(X_{G_{k+1}} = i|X_{G_k}=j) = \mathbb{E}[q_{j\tau i}] \nonumber\\
	&= \sum_{\tau=1}^{M}q_{j\tau i} \P^\pi(\tau|j), \; \forall i,j \in S.
	\end{align}
	\item \textbf{Costs:} In decision epoch $k$, if the state is $j$, then choosing a sampling interval $\tau \in Q$ results in a cost contributing to the average age-penalty which is given by
	\begin{align*}
	c_{j\tau} = \sum_{n=1}^{\tau-1} (\tau - n)(1-p_{jj})p^{n-1}_{jj},
	\end{align*}
	and the cost contributing to the average sampling-interval is given by $\tau$. Note that $c_{j\tau}$ is the expected number of slots the DTMC has spent after moving out of state $j$ in the sampling interval $\tau$. It is easy to see that
	\begin{align}\label{eq:ExpAk+1}
	\mathbb{E}[A_{k+1}|X_{G_k}=j,\tau_k=\tau] = c_{j\tau},\; \forall k \geq 1
	\end{align}
\end{itemize}
\subsection{LP formulations for $\mathcal{P}_1$ and $\mathcal{P}_2$}
We define $z^\pi_{j\tau} = \lim_{k \rightarrow \infty} \P(X_{G_k}=j,\tau_k=\tau)$, the steady-state probability of observing the state-action pair $(j,\tau)$ under a policy $\pi \in \Pi^\text{MR}$. Then, using~\eqref{eq:ExpAk+1}, we obtain
\begin{align*}
\mathbb{E}[A(\pi)] = \sum_{j=1}^{N}\sum_{\tau = 1}^{M} c_{j\tau} z^\pi_{j\tau}\\
\limsup_{K \rightarrow \infty} \frac{\mathbb{E}[\sum_{k=1}^{K}\tau_k]}{K} = \sum_{j=1}^{N}\sum_{\tau = 1}^{M} \tau z^\pi_{j\tau}.
\end{align*}

In the LP formulations for $\mathcal{P}_1$ and $\mathcal{P}_2$, we solve for $z^\pi_{j\tau}$ with the following constraints,
\begin{align}
\sum_{j=1}^{N}\sum_{\tau = 1}^{M} z^\pi_{j\tau} = 1, \label{constraint1}\\
\sum_{\tau = 1}^{M} z^\pi_{i\tau} = \sum_{j=1}^{N}\sum_{\tau = 1}^{M} q_{j\tau i}z^\pi_{j\tau},\; i \in S, \label{constraint2}\\
z^\pi_{j\tau} \geq 0, \; j \in S\text{ and }\tau \in Q. \label{constraint3}
\end{align}
The constraint~\eqref{constraint2} is a consequence of the equilibrium equations for the induced DTMC in the steady state. In the following, we present an equivalent LP formulation for $\mathcal{P}_1$,
\begin{equation}\label{prob3}
\begin{aligned}
& \underset{ \{z^\pi_{j\tau}\} }{\text{minimize}} && \sum_{j=1}^{N}\sum_{\tau = 1}^{M} c_{j\tau} z^\pi_{j\tau}   \\
& \text{s.t.} &&  \sum_{j=1}^{N}\sum_{\tau = 1}^{M} \tau z^\pi_{j\tau} \geq \frac{1}{\nu},\\
&&& \eqref{constraint1}, \eqref{constraint2}, \eqref{constraint3}.
\end{aligned}
\end{equation} 
Let $\{z^*_{j\tau}\}$ denote the optimal solution for~\eqref{prob3}, then the stationary probabilities under $\pi_1^*$ are computed as follows. For $\tau \in Q$,
\begin{align*}
\P^{\pi_1^*}(\tau|j) = \frac{z^*_{j\tau}}{\sum^{M}_{\tau=1} z^*_{i\tau}}, \; j \in S.
\end{align*}
Similarly, an equivalent LP can be formulated for $\mathcal{P}_2$ and $\pi_2^*$ can be obtained.

\subsection{Computing $\nu^\dagger$}
Note that, in $\mathcal{P}_1$ the value of $\nu$ in the constraint can be chosen in the interval $(0,1]$. We are particularly interested in setting $\nu = \nu^\dagger$, because this will give us the minimum achievable average age-penalty for the same sampling frequency achieved by the optimal clairvoyant sampling policy $\pi^\dagger$. We note that $\nu^{\dagger}$ can be obtained by subtracting the percentage of the total frequency of transitions in the DTMC contributed due to self transitions, i.e., transitions from a state to itself, from the total frequency of transitions in the DTMC. Since a transition occurs in every time slot, total frequency of transitions in the DTMC is $1$. The percentage of the total frequency of transitions in the DTMC contributed due to self transitions is given by $\sum_{j=1}^{N} \xi_jp_{jj}$.
The following proposition follows directly from the above analysis.
\begin{proposition}
	Under the optimal clairvoyant sampling policy $\pi^{\dagger}$, the average sampling frequency $\nu^\dagger$ is given by 
	\begin{align*}
	\nu^\dagger = 1 - \sum_{j=1}^{N} \xi_jp_{jj}.
	\end{align*}
\end{proposition}

\begin{figure}[t]
	\centering
	\includegraphics[width = 2.8in]{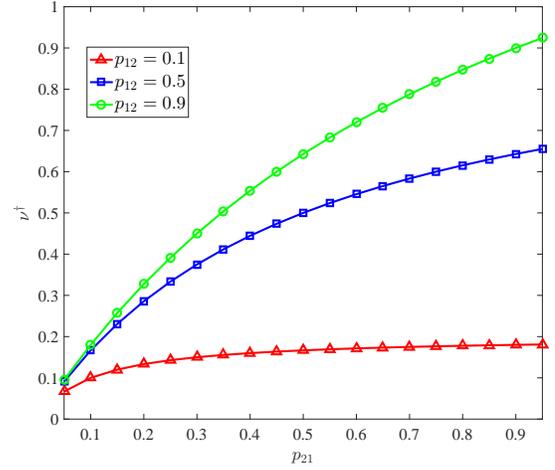}
	\caption{Sampling frequency under $\pi^{\dagger}$ for a two-state Markov chain.}
	\label{fig:nustar_minFreq_20191219}
\end{figure}

For a two-state Markov chain, the steady-state probabilities are given by
\begin{align*}
\xi_1 = \frac{p_{21}}{p_{12}+p_{21}},\quad \text{and} \quad \xi_2 = \frac{p_{12}}{p_{12}+p_{21}},
\end{align*}
and
\begin{align*}
\nu^{\dagger} = \xi_1 p_{12} + \xi_2 p_{21} = \frac{2p_{12}p_{21}}{p_{12}+p_{21}}.
\end{align*}
Figure~\ref{fig:nustar_minFreq_20191219} shows $\nu^\dagger$ versus $p_{21}$ for different values of $p_{12}$.

\section{Numerical Results: Two-State Markov Chain}\label{sec:numerical}
In this section we present numerical analysis for a two-state DTMC. Even though this is the simplest case, it can potentially be used in modelling sources where the set of states can be divided into two sets, for example ``disturbance" vs ``no disturbance", and the events of interest are transitions between these sets. We have implemented the LPs using \textit{linprog} in MATLAB. In the following, we first present two numerical examples to examine the structure of the optimal Markov policies for $\mathcal{P}_1$ and $\mathcal{P}_2$. We then present sampling frequency and age penalty trade-off and a performance comparison between optimal and optimal periodic sampling policies.

\subsubsection*{Example 1}In this example, we solve $\mathcal{P}_1$ for transition probabilities $p_{12} = 0.1$ and $p_{21} = 0.6$, and the constraint on the expected sampling interval is equal to $1/\nu^\dagger = 5.83$. The computation of the optimal policy $\pi_1^*$ results in the following stationary probabilities, 
\begin{align*}
&\P^{\pi_1^*}(\tau = 6|j = 1) = 0.465\text{ and }\P^{\pi_1^*}(\tau = 7|j = 1) = 0.535,\\
&\P^{\pi_1^*}(\tau = 2|j=2) = 1.
\end{align*}
The transition probability out of state $2$ is higher and thus the policy sets $\tau = 2$ when the observed state is $2$. The minimum expected age penalty is computed to be $1.416$.
An optimal periodic sampling policy chooses $\tau = \tau^\dagger = \lceil 1/\nu^\dagger \rceil = 6$.

\subsubsection*{Example 2} In this example, we solve $\mathcal{{P}}_2$ when $p_{12} = 0.9$, $p_{21} = 0.9$, and the expected age penalty is upper bounded by $d = 1$. The computation of the optimal policy $\pi_2^*$ results in the following stationary probabilities. 
\begin{align*}
\P^{\pi_2^*}(\tau = 2|j) = 0.899\text{ and }\P^{\pi_2^*}(\tau = 3|j) = 0.101\text{ for }j = 1,2.
\end{align*}
The minimum expected sampling frequency is computed to be $0.476$. The optimal periodic sampling policy chooses $\tau = d+1 = 2$, and hence its sampling frequency is $0.5$.

\subsection{Performance Comparison}
\begin{figure}[t]
	\centering
	\includegraphics[width = 2.8in]{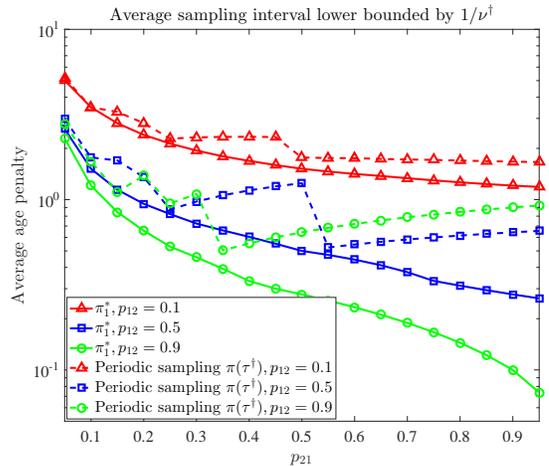}
	\caption{Average age penalties achieved by $\pi^*_1$ and the optimal periodic sampler for different $p_{12}$ and $p_{21}$ values.}
	\label{fig:minAge_periodicAge_20191219}
\end{figure}

In Figure~\ref{fig:minAge_periodicAge_20191219}, we compare the average age penalties achieved by optimal periodic sampler and the optimal policy $\pi^*_1$ obtained by solving $\mathcal{P}_1$ under the constraint $\nu = \nu^{\dagger}$. Recall that for this case, the optimal periodic sampler sets the sampling interval equal to $\tau^{\dagger} = \lceil 1/\nu^{\dagger} \rceil$. From the figure, we observe that for lower transition probabilities between the sates, i.e., lower $p_{12}$ and $p_{21}$ values, periodic sampler achieves age penalties only slightly higher than that of the optimal policy, because in this case the optimal policy is also choosing sampling intervals close to that of the periodic sampler. The gap between them, however, increases significantly for higher transition probabilities. The zigzag pattern of the periodic sampler can be attributed to the ceil function used in computing the sampling interval.\\

\begin{figure}[t]
	\centering
	\includegraphics[width = 2.8in]{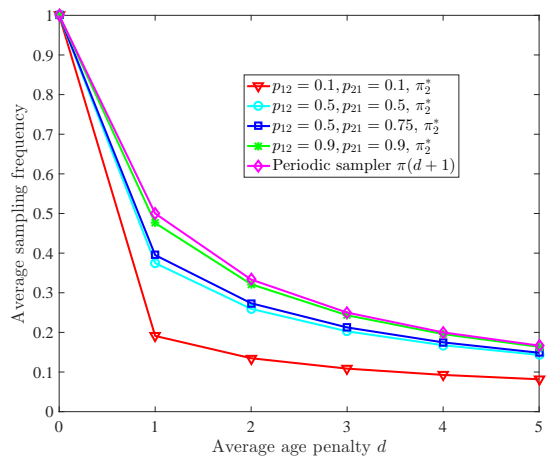}
	\caption{Sampling frequency vs age penalty trade-off: expected sampling frequencies achieved by $\pi^*_2$ and the optimal periodic sampler by varying age penalty constraint value.}
	\label{fig:freq_vs_age}
\end{figure}

\begin{figure}[t]
	\centering
	\includegraphics[width = 2.8in]{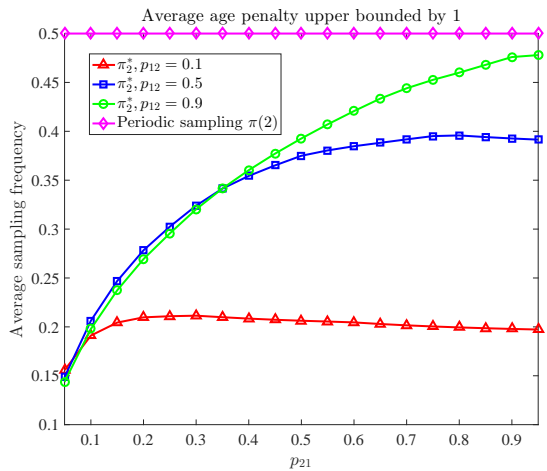}
	\caption{Average sampling frequencies achieved by $\pi^*_2$ and the optimal periodic sampler for varying $p_{12}$, and $d = 1$.}
	\label{fig:minFreq_AgeLimit1_20191217}
\end{figure}

In Figures~\ref{fig:freq_vs_age}, and~\ref{fig:minFreq_AgeLimit1_20191217} we compare average sampling frequencies achieved by the optimal periodic sampler and the optimal policy $\pi^*_2$ by solving $\mathcal{P}_2$. From Figure~\ref{fig:freq_vs_age}, we observe the trade-off between achievable sampling frequencies and age penalties. As expected, for age penalty constraint of one time slot, i.e. $d = 1$, the achievable sampling frequency is lower than $0.5$ for both policies. However, $\pi^*_2$ results in much lower sampling frequencies for lower transition probabilities. In Figure~\ref{fig:minFreq_AgeLimit1_20191217}, we set $d = 1$ and thus the optimal periodic sampler samples every $2$ time slots with sampling frequency $0.5$. On the other hand, $\pi^*_2$ provides much lower sampling frequencies when either of the transition probabilities are small.

\begin{figure}[t]
	\centering
	\includegraphics[width = 2.8in]{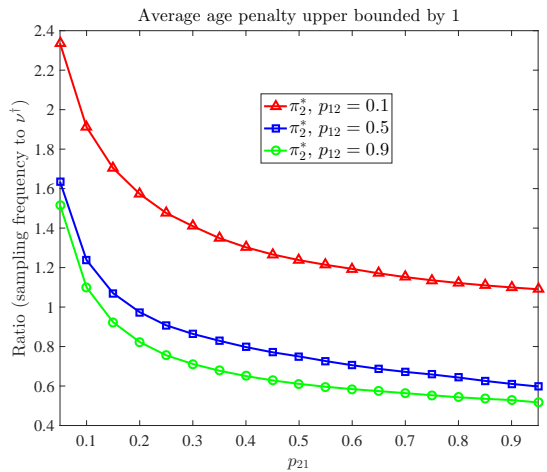}
	\caption{Ratio between the expected sampling frequency achieved by $\pi^*_2$ and $\nu^\dagger$.}
	\label{fig:ratio_20191219}
\end{figure}

Finally, in Figure~\ref{fig:ratio_20191219}, we present the ratio between the expected sampling frequency achieved by $\pi^*_2$ and $\nu^\dagger$, under average age penalty constraint $d=1$. We note  that under $\pi^{\dagger}$ the age penalty is always zero. This cannot be achieved by any causal policy with a sampling frequency strictly less than one. Nonetheless, an interesting observation from the figure is that by allowing a small age penalty $d = 1$, the optimal policy $\pi^*_2$ can achieve lower sampling frequency than $\nu^{\dagger}$ when transition probabilities are higher, say $p_{12} = 0.9$ and $p_{21} = 0.9$. For lower transition probabilities $p_{12} = 0.1$ and $p_{21} = 0.1$, the ratio is always greater than $1$, i.e., optimal policy $\pi^*_2$ couldn't achieve the sampling frequency  $\nu^{\dagger}$ and may require more relaxation in the age penalty constraint. In conclusion, for lower transition probabilities, i.e., if the events become rare, the optimal policy performs worse with respect optimal clairvoyant sampling policy.

\section{Related Works}\label{sec:related}
In the AoI literature, the works \cite{Kam2018,SunBenjamin2018,Feng2020,Inoue2019} considered remote monitoring/estimation of the states of a Markov source. In \cite{Kam2018}, the authors studied remote state estimation of a two-state Markov Chain where the communication delay is geometrically distributed. They computed average AoI and estimation error for two sampling policies: zero-wait policy, which generates a sample when the channel is idle, and sample-at-change policy, which generates a sample when the channel is idle and a  transition to a state different from the previous sample occurs. The authors in~\cite{SunBenjamin2018} proposed a freshness metric based on the mutual information between the current state of the source and the received states at a remote monitor, and solved an optimal sampling problem for maximizing the mutual information.  In~\cite{Feng2020}, the authors analysed freshness by proposing a closely related metric based on conditional entropy, where current state and the states in the past till the time of generation of the freshest sample at the monitor are conditioned with respect to this freshest sample. Displaying the state of a continuous-time Markov chain source at a remote monitor was studied in~\cite{Inoue2019}. The authors analysed the probability of error in displaying the correct state of the source. In our system model, we consider staleness only at the sampler. The age penalty metric we studied is different from the above works and is used to uniquely capture the trade-off between stateless and sampling frequency by considering the dynamics of the Markov chain. 

The problem of when to sample next has been studied for many years in control theory, see for example~\cite{Kushner1964,Skafidas1998,Rabi2012,Ross2015}. In~\cite{Kushner1964} (\cite{Skafidas1998}), the authors considered the off-line (on-line) problem of choosing the time instants to sample sensor measurements to minimize a Linear Quadratic Gaussian (LQG) cost in a Linear Time Invariant (LTI) system. In~\cite{Rabi2012}, the authors considered minimizing squared error distortion for state estimation of a Markov source under a constraint on maximum number of transmitted samples. We note however that, in this work the sensor is assumed to samples the process continuously but only transmits certain samples based on some criterion (event-triggering). In~\cite{Ross2015}, the authors studied the design of sampling intervals such that the stability of a non-linear stochastic dynamical system is ensured. In all the above works, the objective is either to minimize estimation error or control cost or ensure stability of the system.

Perhaps the most relevant application of the problem we have studied is the web crawling application~\cite{Cho2003,Olsten2010}. The authors in~\cite{Cho2003} have solved a static optimization problem for computing optimal fixed intervals between downloads for different web pages. To the best of our knowledge, dynamic policies that use the state of the system have not been studied in this line of work; see~\cite{Olsten2010} for a survey. In contrast to the above works, we considered the set of causal sampling policies and studied the trade-off between sampling frequency and age penalty for detecting state transitions in a finite-state DTMC.

\section{Conclusion}\label{sec:conclusion}
We have studied the trade-off between sampling frequency and staleness for detecting transitions of a DTMC to new states. The staleness of the $k$th sample is quantified using age penalty, which is defined as the time elapsed since the first transition out of the state in the $k-1$ sample. The formulated problems $\mathcal{P}_1$ and $\mathcal{P}_2$ are CMDPs and were solved by deriving equivalent LPs. We have provided a closed-form expression for $\nu^\dagger$, the sampling frequency under the optimal clairvoyant sampling policy. Even though our problem setting looked simple, the numerical examples revealed that the optimal policies have randomized Markov policy structure, i.e., simple deterministic optimal policies may not exist for this problem. Apart from the superior performance of the computed optimal policy over optimal periodic sampling policy, we found that by allowing a small age penalty the optimal policy achieves sampling frequency lower than $\nu^\dagger$ in some cases.

We leave comprehensive simulation results considering $N>2$ for future work. We would like to explore different age penalties and study the trade-off when there are multiple sources. Finally, we are interested in studying the problem for different models for the information source. 


\end{document}